\documentclass[american,twocolumn,nofootinbib]{revtex4-2}
\usepackage[T1]{fontenc}
\usepackage[utf8]{inputenc}
\usepackage{color}
\usepackage{babel}
\usepackage{amsmath}
\usepackage{amssymb}
\usepackage{graphicx}
\usepackage[pdfusetitle,
 bookmarks=true,bookmarksnumbered=false,bookmarksopen=false,
 breaklinks=false,pdfborder={0 0 1},backref=false,colorlinks=true]
 {hyperref}



\begin{document}
\title{Transverse momentum dependence of $\Omega/\phi$ ratio in high energy collisions}
\author{Hai-hong Li}
\affiliation{School of Physics and Electronic Engineering, Jining University, Shandong 273155, China}
\author{Jun Song }
\affiliation{School of Physics and Electronic Engineering, Jining University, Shandong 273155, China}
\author{Feng-lan Shao}
\affiliation{School of Physics and Physical Engineering, Qufu Normal University, Shandong 273165, China}
\begin{abstract}
We apply a constituent quark equal-velocity combination model to study the $p_{T}$ dependence of $\Omega/\phi$ ratio in $pp$, $p$-Pb and Pb-Pb collisions at LHC energies. We demonstrate that the relative change rate of the $\Omega/\phi$ ratio is dominated by a discrete curvature property of $p_{T}$ spectrum of strange quarks just before hadronization. Using experimental data of $\phi$ mesons after a quark number scaling operation, we extract $p_{T}$ spectrum of strange quarks just before hadronization and study its curvature property in high and low multiplicity events in $pp$ collisions at $\sqrt{s}=$ 7, 13 TeV, $p$-Pb collisions at $\sqrt{s_{NN}}$= 5.02 TeV and Pb-Pb collisions at $\sqrt{s_{NN}}$= 2.76 TeV. We apply these curvature properties of $p_{T}$ spectra of strange quarks to explain the observed $p_{T}$ dependence of $\Omega/\phi$ ratio in those collisions. We discuss the possible origin of the change of curvature property for $p_{T}$ spectra of strange quarks just before hadronization by considering the influence of strong collective flow formed in partonic stage evolution in collisions at LHC energies. 
\end{abstract}
\maketitle

\section{Introduction}

The ratio of baryon to meson as the function of transverse momentum ($p_{T}$) is a sensitive physical quantity to the hadronization mechanism of final-state parton systems created in high energy collisions. Measurements for the enhancement of $p/\pi$ ratio at RHIC \citep{PHENIX:2001vgc,STAR:2006uve} inspire the application of quark combination mechanism to describe the hadronization of the hot dense quark matter created in ultra-relativistic heavy-ion collisions \citep{Hwa:2002tu,Hwa:2006vb,Greco:2003xt,Chen:2006vc}.  $p/\pi$ ratio and $\Lambda/K_{S}^{0}$ ratio are two most easily measured baryon/meson ratios which were widely reported in $pp$, $p$A and AA collisions at RHIC and LHC energies \citep{STAR:2006nmo,ALICE:2013cdo,ALICE:2016dei,ALICE:2014juv,ALICE:2013wgn,STAR:2006pcq,ALICE:2020jsh,STAR:2006uve,STAR:2019bjj}, and widely studied and discussed in various theoretical models of hadron production \citep{Hwa:2002tu,Greco:2003xt,Chen:2006vc,Hwa:2006vb,Werner:2012sv,Pierog:2013ria,Begun:2014rsa,Minissale:2015zwa,Bierlich:2015rha,JETSCAPE:2025wjn}.  In the last few years, baryon/meson ratios in charm sector such as $\Lambda_{c}^{0}/D^{0}$ are also measured \citep{ALICE:2017thy,LHCb:2018weo,LHCb:2022ddg,ALICE:2022exq,CMS:2023frs}, which sheds new light on the hadronization of charm quark in the low and intermediate $p_{T}$ range in high energy collisions \citep{Cho:2019lxb,Minissale:2020bif,Song:2023nzu,Zhao:2023nrz,Altmann:2024kwx}. 

Because of only involving strange (anti-)quarks, the productions of $\Omega$ and $\phi$ are particularly interesting to reveal the property of hadron production at the hadronization of soft quark systems created in high energy collisions. As we known, strange hadrons have relatively small interaction cross-section with other hadrons and suffer weakly the influence of hadronic re-scatterings \citep{Shor:1984ui,vanHecke:1998yu}.  In comparison with $\pi$, proton, $K_{S}^{0}$, the decay influence to $\Omega$ and $\phi$ is also relatively weak. These advantages make the $\Omega$ and $\phi$ are clean probes of hadron production mechanism at hadronization stage. However, $\Omega/\phi$ ratio is not often reported in experiments \citep{STAR:2008bgi,ALICE:2014jbq} mainly because of the relatively poor statistics for $\Omega$ baryon and theoretical studies in various popular combination(coalescence) models are also not many \citep{Chen:2006vc,Proceedings:2007ctk,Hwa:2006vb,Jin:2018lbk,Pu:2018eei}. 

\begin{figure}[h]
\centering{}\includegraphics[width=0.9\linewidth]{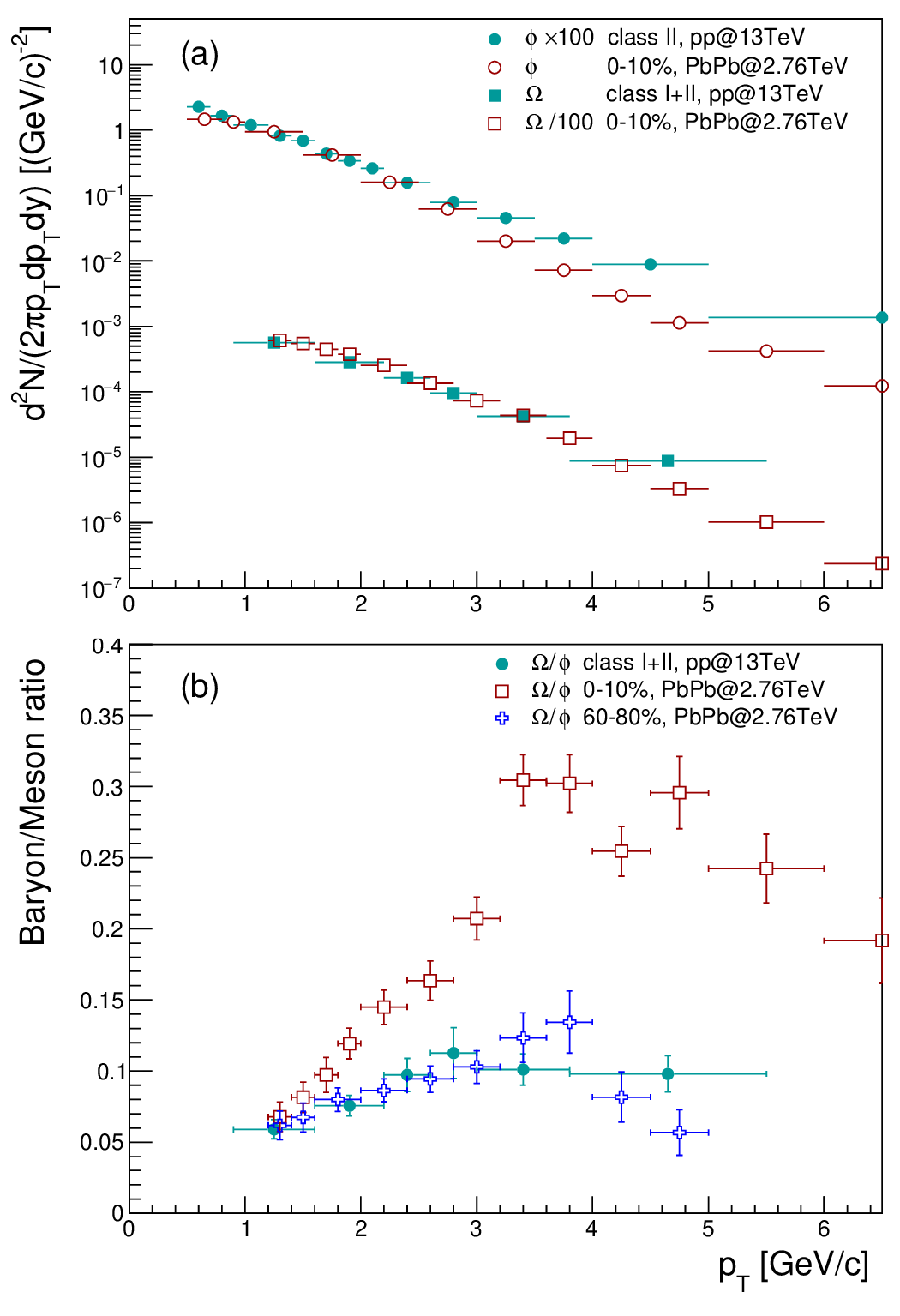}\caption{(a) $p_{T}$ spectra of $\Omega$ and $\phi$ in $pp$ collisions
at $\sqrt{s}=13$ TeV and in Pb-Pb collisions at $\sqrt{s_{NN}}=$2.76
TeV and (b) the $\Omega/\phi$ ratio in the two collision systems.
Data are taken from \citep{ALICE:2019avo,ALICE:2013xmt,ALICE:2017ban,ALICE:2019etb}.
}\label{fig: Omg_phi_ratio_compare}
\end{figure}

In the last few years, LHC has accumulated rich data of $\Omega$ and $\phi$ in $pp$, $p$-Pb and Pb-Pb collisions which makes the analysis of $\Omega/\phi$ ratio plausible. In order to see the underlying dynamics in the $\Omega/\phi$ ratio as the function of $p_{T}$, we select two collision systems, i.e., high multiplicity $pp$ collisions at $\sqrt{s}=$ 13 TeV and central Pb-Pb collisions at $\sqrt{s_{NN}}=$ 2.76 TeV. In Fig.~\ref{fig: Omg_phi_ratio_compare}(a), we show the experimental data for $p_{T}$ spectra of $\phi$ and $\Omega$ (i.e.  $\Omega^{-}+\bar{\Omega}^{+}$) in the event class I+II in $pp$ collisions at $\sqrt{s}=$13 TeV \citep{ALICE:2019avo,ALICE:2019etb} and in 0-10\% centrality in Pb-Pb collisions at $\sqrt{s_{NN}}=$2.76 TeV \citep{ALICE:2013xmt,ALICE:2017ban}. Two collision systems are quite different in system volume, which can be inferred from the charged-particle multiplicity at mid-rapidity $\left\langle dN_{ch}/d\eta\right\rangle =21.0\pm0.25$
in class I+II $pp$ collisions at $\sqrt{s}=$13 TeV \citep{ALICE:2019avo} and $\left\langle dN_{ch}/d\eta\right\rangle =1446\pm39$ in 0-10\% centrality in Pb-Pb collisions at $\sqrt{s_{NN}}=$ 2.76 TeV \citep{ALICE:2010mlf}.  However, we see in Fig.~\ref{fig: Omg_phi_ratio_compare}(a) that, except the global normalization (i.e., yield density at mid-rapidity), the inclusive $p_{T}$ spectra of $\Omega$ or $\phi$ in $pp$ collisions are similar to those in Pb-Pb collisions in the low $p_{T}$ range but they are flatter than the latter in the intermediate $p_{T}$ range. The averaged transverse momentum $\left\langle p_{T}\right\rangle $ for $\phi$ in $pp$ collisions is $1.456\pm0.015$ GeV/$c$ \citep{ALICE:2019avo} which is also larger than $1.325\pm0.028$ GeV/$c$ in central Pb-Pb collisions \citep{ALICE:2017ban}.  Using above data of $\Omega$ and $\phi$, we evaluate the $\Omega/\phi$
ratio \footnote{The bin of $p_{T}$ spectra of $\Omega$ is usually different from that of $\phi$. Considering that data points of $\phi$ are richer than those of $\Omega$, we firstly interpolate the density $dN/dp_{T}$ of $\phi$ for the selected $p_{T}$ bin from $\Omega$ data and then evaluate the ratio $\Omega/\phi$. In interpolation, we take three data points of $\phi$ that completely cover the selected bin of $\Omega$ and fit these three data points with function $\exp\left[P_{2}(p_{T})\right]$ where $P_{2}(p_{T})$ is the second polynomial. }and show results in Fig.~\ref{fig: Omg_phi_ratio_compare}(b). In the small $p_{T}$ range $\Omega/\phi$ ratio in $pp$ collisions is close to (only slightly smaller than) that in central Pb-Pb collisions but in the range $2\lesssim p_{T}\lesssim5$ GeV/$c$ it is obviously smaller than the latter. Actually, the $\Omega/\phi$ ratio in high-multiplicity $pp$ collisions is quite similar to that in peripheral (60-80\% centrality) Pb-Pb collisions at $\sqrt{s_{NN}}=$2.76 TeV. In other words, although we observe the flatter $p_{T}$ spectra of $\Omega$ or $\phi$ in $pp$ collisions, which means the relatively more $\Omega$ or $\phi$ at intermediate $p_{T}$ in $pp$ collisions, the significant enhancement of $\Omega$ production in comparison with $\phi$ does not happen in high-multiplicity $pp$ collisions.

\begin{figure}[h]
\centering{}\includegraphics[width=0.9\linewidth]{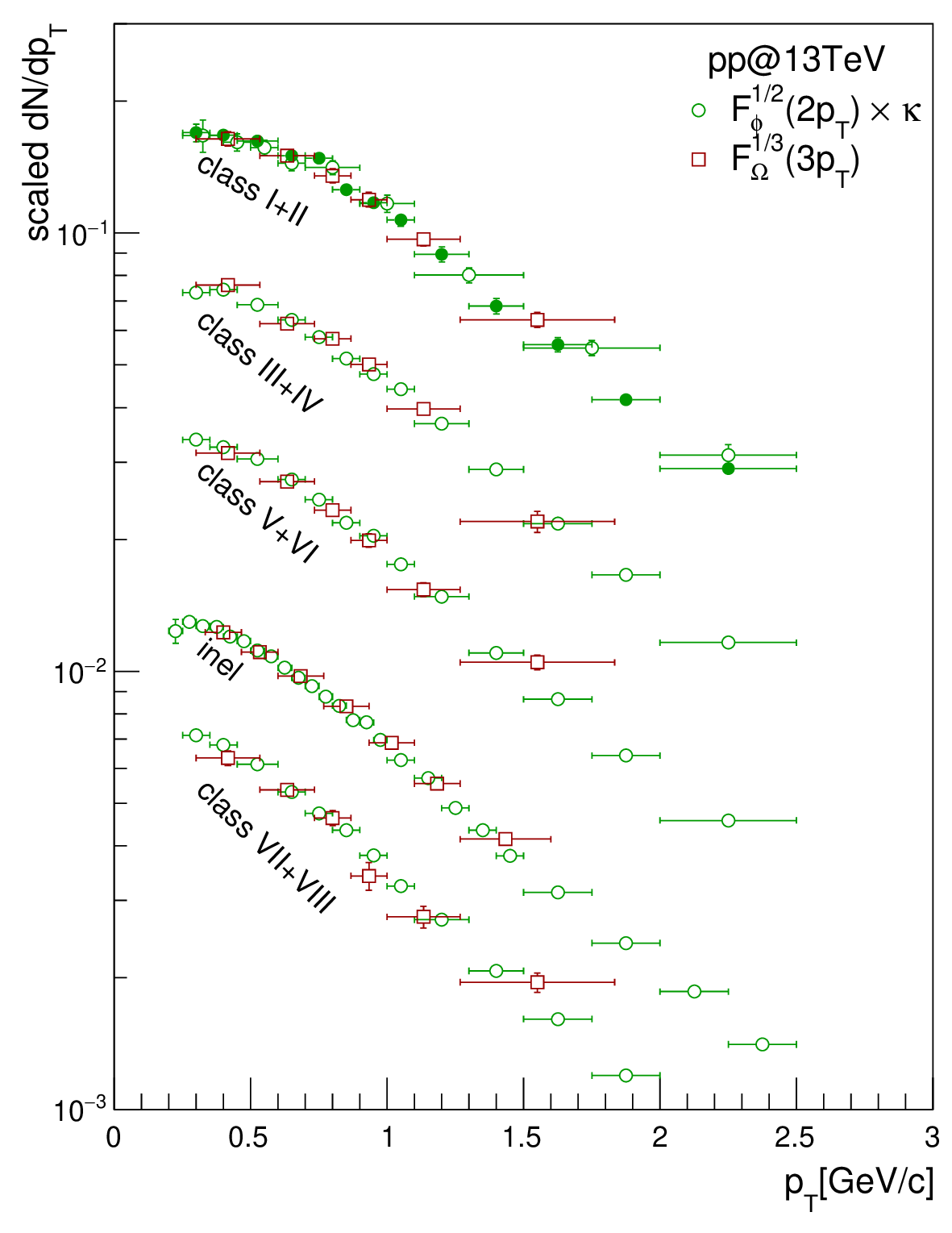}
    \caption{The quark number scaling test for $p_{T}$ spectra of $\Omega$ and
$\phi$ in $pp$ collisions at $\sqrt{s}=13$ TeV. Data are taken
from \citep{ALICE:2019avo,ALICE:2013xmt,ALICE:2017ban,ALICE:2019etb,ALICE:2020jsh}.
}\label{fig: Omg_phi_qns_pp13TeV}
\end{figure}

An interesting question is what the underlying dynamics is responsible for such significant change of $\Omega/\phi$ ratio from $pp$ collisions to central Pb-Pb collisions at LHC energies. In previous studies of $p/\pi$ ratio and $\Lambda/K_{s}^{0}$ ratio, it is usually argued that the change of hadron production mechanism from parton fragmentation to quark combination is a sound explanation for such a significant change of baryon/meson ratios at intermediate $p_{T}$ in small and large collision systems. However, our recent studies show that quark combination mechanism is also quite effective to explain the $p_{T}$ spectra of hadrons in the low and intermediate $p_{T}$ range in $pp$ collisions at LHC energies \citep{Gou:2017foe,Zhang:2018vyr,Li:2021nhq}.  A typical evidence is the quark number scaling property for $p_{T}$ spectra of $\Omega$ and $\phi$ \citep{Song:2017gcz,Gou:2017foe,Zhang:2018vyr,Li:2021nhq}
\begin{equation}
F_{\Omega}^{1/3}(3p_{T})=\kappa F_{\phi}^{1/2}(2p_{T})\label{eq:qns_Omg_phi}
\end{equation}
where $F(p_{T})\equiv dN/dp_{T}$ and $\kappa$ is the $p_{T}$ independent coefficient.  In Fig.~\ref{fig: Omg_phi_qns_pp13TeV}, we show the test of this scaling property for experimental data of $\Omega$ and $\phi$ at mid-rapidity in different multiplicity classes in $pp$ collisions at $\sqrt{s}=13$ TeV \citep{ALICE:2019avo,ALICE:2013xmt,ALICE:2017ban,ALICE:2019etb,ALICE:2020jsh}.  Within the current statistics, we see that the scaling property holds well in high multiplicity events as well as in inelastic events. The scaling property for data of Pb-Pb collisions at $\sqrt{s_{NN}}=$2.76 TeV was already tested in \citep{Song:2019sez}. Since this scaling property can be understood through the quark equal-velocity combination mechanism by $F_{\Omega}(3p_{T})\propto F_{s}^{3}(p_{T})$ and $F_{\phi}(2p_{T})\propto F_{s}^{2}(p_{T})$ with $F_{s}(p_{T})$ being interpreted as $p_{T}$ spectra of strange quarks just before hadronization. Therefore, we argue that the significant change of $\Omega/\phi$ ratio from $pp$ collisions to central Pb-Pb collisions at LHC energies might not be because of the change of hadronization mechanism but be probably due to other dynamics such as the change in property of quark $p_{T}$ spectra in different collisions systems, which is the study focus of this paper. 

In this paper, we apply a constituent quark equal-velocity combination mechanism to study the production of $\Omega$ and $\phi$ in high energy $pp$, $p$-Pb and Pb-Pb collisions. We explain the underlying dynamics that influences the $p_{T}$ dependence of $\Omega/\phi$ ratio by particularly focusing on the shape property of $p_{T}$ spectrum of strange quarks just before hadronization. We propose a discrete curvature for $p_{T}$ spectrum of strange quarks which can directly determine the increase or decrease rate of $\Omega/\phi$ ratio as the function of $p_{T}$. We analyze this curvature property of strange quark $p_{T}$ spectra through the scaled data of $\phi$ in $pp$, $p$-Pb and Pb-Pb collisions at LHC energies and apply it to explain the data of $\Omega/\phi$ ratios in these collisions. 

The paper is organized as follows. In Sec. \ref{sec:evc_model}, we apply an equal-velocity quark combination mechanism to demonstrate how the curvature property of $p_{T}$ spectrum strange quarks influences the $p_{T}$ dependence of $\Omega/\phi$ ratio. In Sec. \ref{sec:fs_Omg_phi_ratio}, we discuss the curvature property of strange quark $p_{T}$ spectra obtained from the scaled data of $\phi$ in $pp$, $p$-Pb and Pb-Pb collisions at LHC energies and explain data of $\Omega/\phi$ ratios in these collisions. Sec. \ref{sec:summary} gives the summary and discussion. 

\section{$p_{T}$ dependence of $\Omega/\phi$ ratio in EVC model}\label{sec:evc_model}

We apply a particular quark equal-velocity combination (EVC) model \citep{Song:2017gcz,Gou:2017foe} to study the $p_{T}$ dependence of $\Omega/\phi$ ratio in high energy collisions. In order to model the hadronization of final-state parton system created in high energy collisions, we take constituent quarks and anti-quarks as the effective degrees of freedom of the parton system just before hadronization and assume that the formation of the hadron is mainly characterized by the combination of these constituent quarks and anti-quarks with the same velocity. In this case, the $p_{T}$ distribution of a hadron ($F(p_{T})\equiv dN/dp_{T}$) is the product of those of (anti-)quarks
\begin{align}
F_{B_{i}}\left(p_{T}\right) & =\kappa_{B_{i}}F_{q_{1}}\left(x_{1}p_{T}\right)F_{q_{2}}\left(x_{2}p_{T}\right)F_{q_{3}}\left(x_{3}p_{T}\right),\label{eq:fbi}\\
F_{M_{i}}\left(p_{T}\right) & =\kappa_{M_{i}}F_{q_{1}}\left(x_{1}p_{T}\right)F_{\bar{q}_{2}}\left(x_{2}p_{T}\right).\label{eq:fmi}
\end{align}
Here, moment fractions satisfy $x_{1}+x_{2}+x_{3}=1$ with $x_{i}=m_{i}/(m_{1}+m_{2}+m_{3})$ ($i=1,2,3)$ in baryon formation and $x_{1}+x_{2}=1$ with $x_{i}=m_{i}/(m_{1}+m_{2})$ ($i=1,2)$ in meson formation. $m_{i}$ is constituent mass of quark $q_{i}$. Coefficients $\kappa_{B_{i}}$ and $\kappa_{M_{i}}$ are independent of $p_{T}$ and can be further decomposed and determined by the unitarity of hadronization and few experimental constraints \citep{Gou:2017foe,Song:2020kak}. 

Applying EVC to $\Omega^{-}$ and $\phi$, we obtain 
\begin{align}
F_{\Omega^{-}}(p_{T}) & =\kappa_{\Omega^{-}}F_{s}^{3}(p_{T}/3),\\
F_{\phi}(p_{T}) & =\kappa_{\phi}F_{s}^{2}(p_{T}/2).
\end{align}
Considering LHC case where the productions of $\Omega^{-}$ and $\bar{\Omega}^{+}$
are almost symmetric, $\kappa$ coefficient of $\Omega$ (i.e., $\Omega^{-}+\bar{\Omega}^{+}$)
and $\phi$ can be written as 
\begin{align}
\kappa_{\Omega} & =\frac{2}{3a}A_{sss}/N_{q}^{2},\\
\kappa_{\phi} & =\left(1-\frac{1}{a}\right)A_{ss}C_{\phi}/N_{q}.
\end{align}
Here, $N_{q}$ is the number of all quarks in the system at hadronization and $N_{q}=N_{u}+N_{d}+N_{s}$ in the case that only light-flavor quarks are considered. $a$ is a parameter that describes the production competition between baryon and meson by the relation $N_{M}/N_{B}=3(a-1)$ in quark combination model \citep{Song:2013isa}. It is relatively stable for light-flavor hadron production and is about $4.86\pm0.14$ according to our previous study \citep{Shao:2017eok}. $C_{\phi}$ is the condition probability to form a $\phi$ meson from a $s\bar{s}$ pair that is destined to form a meson. $C_{\phi}$ is about $0.33\pm0.02$ by fitting data of $\phi/K$ yield ratio \citep{ALICE:2017ban,ALICE:2019etb}.  $A_{ss}^{-1}=2\int dp_{T}[F_{s}^{(n)}(p_{T})]^{2}$ and $A_{sss}^{-1}=3\int dp_{T}[F_{s}^{(n)}(p_{T})]^{3}$ are determined by the integration of the square and cubic of the normalized $p_{T}$ distribution of strange quarks $F_{s}^{(n)}(p_{T})\equiv F_{s}(p_{T})/N_{s}$, respectively. 

The $\Omega/\phi$ ratio in EVC model can be written as 
\begin{equation}
\frac{F_{\Omega}(p_{T})}{F_{\phi}(p_{T})}=\frac{2\lambda_{s}}{3C_{\phi}\left(a-1\right)(2+\lambda_{s})}\frac{A_{sss}F_{s}^{(n)}(p_{T}/3)}{A_{ss}F_{s}^{(n)}(p_{T}/2)}.
\end{equation}
The second fraction in the right hand side of the equation corresponds to the normalized distribution of $\Omega$ and $\phi$ which gives the $p_{T}$ dependence of the $\Omega/\phi$ ratio. The first fraction denotes the global yield ratio of $\Omega$ to $\phi$. $\lambda_{s}=N_{s}/N_{u}\approx N_{s}/N_{d}$ is a factor describing the production suppression of strange quarks relative to up/down quarks where $N_{u}\approx N_{d}$ is often assumed in collisions at LHC energies. 

In order to reveal the key ingredient that mostly influences the $p_{T}$ dependence of the $\Omega/\phi$ ratio, we evaluate the slope of the ratio 
\begin{equation}
\left[\frac{F_{\Omega}(p_{T})}{F_{\phi}(p_{T})}\right]^{'}=\frac{F_{\Omega}(p_{T})}{F_{\phi}(p_{T})}\left[\frac{\partial\ln F_{s}^{(n)}(\frac{p_{T}}{3})}{\partial\left(p_{T}/3\right)}-\frac{\partial\ln F_{s}^{(n)}(\frac{p_{T}}{2})}{\partial\left(p_{T}/2\right)}\right]
\end{equation}
and obtain the relative change rate of the ratio 
\begin{equation}
\left[\ln\frac{F_{\Omega}(p_{T})}{F_{\phi}(p_{T})}\right]^{'}=\left[\frac{\partial\ln F_{s}^{(n)}(\frac{p_{T}}{3})}{\partial\left(p_{T}/3\right)}-\frac{\partial\ln F_{s}^{(n)}(\frac{p_{T}}{2})}{\partial\left(p_{T}/2\right)}\right].
\end{equation}
It is only dependent on the shape property of strange quarks. Using the mean-value theorem 
\begin{equation}
\frac{\partial\ln F_{s}^{(n)}(\frac{p_{T}}{3})}{\partial\left(p_{T}/3\right)}-\frac{\partial\ln F_{s}^{(n)}(\frac{p_{T}}{2})}{\partial\left(p_{T}/2\right)}=-\frac{p_{T}}{6}\left[\ln F_{s}^{(n)}(\xi)\right]^{''}
\end{equation}
where $p_{T}/3\leq\xi\leq p_{T}/2$, we obtain that the curvature (convex or concave) property of $F_{s}(p_{T})$ is the key ingredient to influence the $p_{T}$ dependency of $\Omega/\phi$ ratio.

\section{curvature property of $p_{T}$ spectra of strange quark and its influence on $\Omega/\phi$ ratio }\label{sec:fs_Omg_phi_ratio}

\begin{figure}[h]

\centering{}\includegraphics[width=0.9\columnwidth]{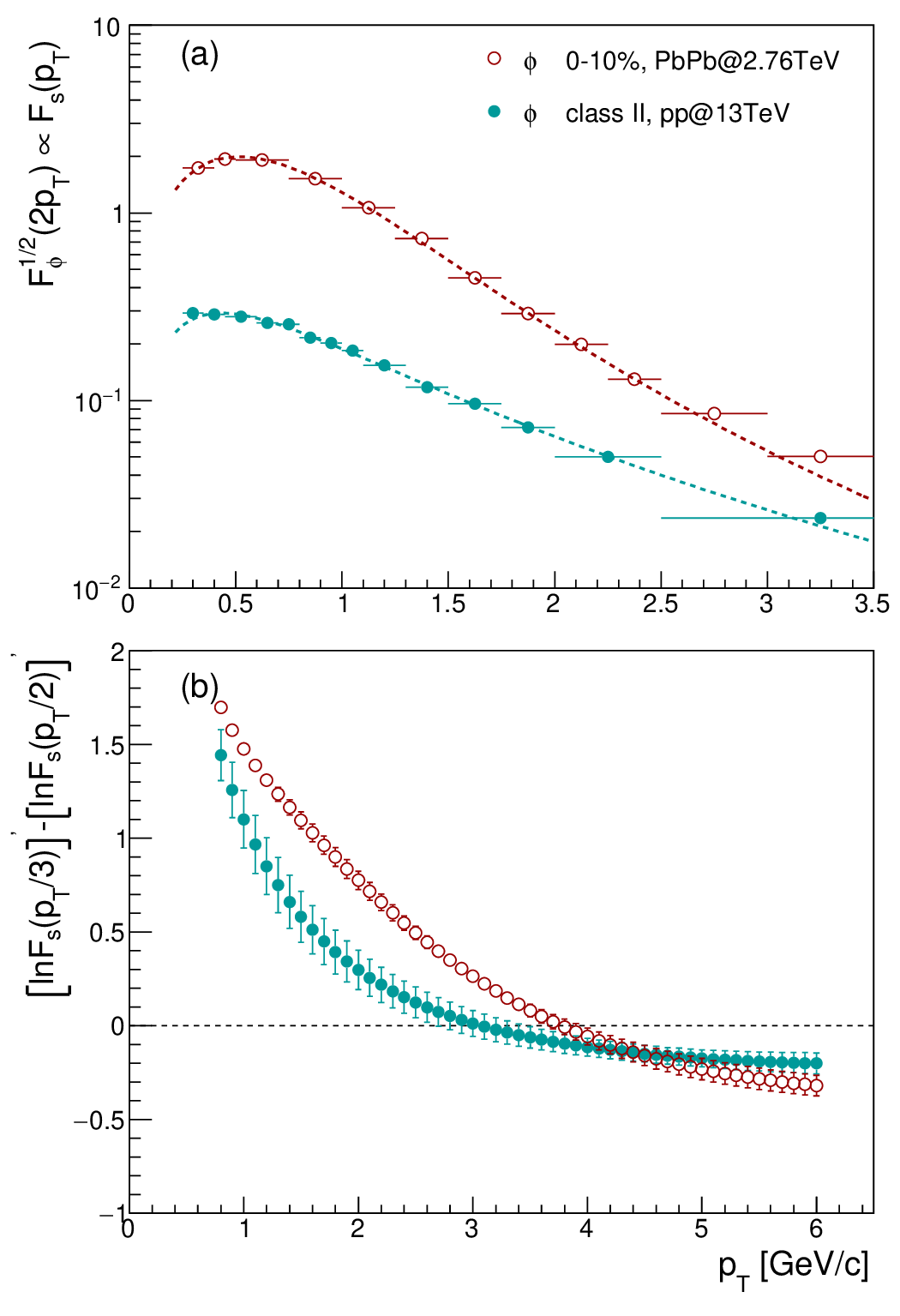}
    \caption{(a) The scaled $p_{T}$ spectra of $\phi$ in high-multiplicity (class II) $pp$ collisions at $\sqrt{s}=$ 13 TeV and in central Pb-Pb collisions at $\sqrt{s_{NN}}=$ 2.76 TeV. Lines are fittings of Levy-Tsallis function; (b) the curvature $\left[\ln F_{s}(p_{T}/3)\right]^{'}-\left[\ln F_{s}(p_{T}/2)\right]^{'}$ of $p_{T}$ spectra of strange quarks. }\label{fig: fs_curvature}
\end{figure}

Considering that experimental data of $\phi$ are usually richer than those of $\Omega,$ we can extract $F_{s}(p_{T})$ by experimental data of $\phi$ 
\begin{equation}
F_{s}(p_{T})=\frac{1}{\sqrt{\kappa_{\phi}}}\left[F_{\phi}^{(data)}(2p_{T})\right]^{1/2}.
\end{equation}
In Fig.~\ref{fig: fs_curvature}(a), we show $p_{T}$ spectra of strange quarks by scaling data of $\phi$ in high-multiplicity (class II) $pp$ collisions at $\sqrt{s}=$ 13 TeV and in central Pb-Pb collisions at $\sqrt{s_{NN}}=$ 2.76 TeV \citep{ALICE:2019etb,ALICE:2017ban}. 

In order to evaluate the curvature of strange quark $p_{T}$ spectrum, we fit discrete data points of $F_{s}(p_{T})$ by the Levy-Tsallis function \citep{Tsallis:1987eu} and its two variants 
\begin{align}
\frac{dN^{Levy}}{2\pi p_{T}dp_{T}} & =N\left[1+\frac{m_{T}-m}{nc}\right]^{-n},\label{eq:fpt_levy_v0}\\
\frac{dN^{Levy-v2}}{2\pi p_{T}dp_{T}} & =N\left[1+\frac{\left(m_{T}-m\right)^{a}}{nc}\right]^{-n},\label{eq:fpt_levy_v2}\\
\frac{dN^{Levy-v3}}{2\pi p_{T}dp_{T}} & =N\left[1+\frac{\left(m_{T}-m-p_{T}v\right)}{nc\sqrt{1-v^{2}}}\right]^{-n}.\label{eq:fpt_levy_v3}
\end{align}
where $N$, $m$, $n$, $c$ are parameters of Levy-Tsallis function and parameters $a$ and $v$ are added in the latter two functions for better tuning the shape. The purpose of choosing three fitting functions is to reduce the bias from the selection of fitting function.  In practical fittings, the latter two functions are usually slightly better than the standard Levy-Tsallis function. The final result of the curvature $\left[\ln F_{s}(p_{T}/3)\right]^{'}-\left[\ln F_{s}(p_{T}/2)\right]^{'}$ is the average of three functions weighted by the inverse of their fitting quality values $\chi^{2}$. 

In Fig.~\ref{fig: fs_curvature}(b), we show the curvature $\left[\ln F_{s}(p_{T}/3)\right]^{'}-\left[\ln F_{s}(p_{T}/2)\right]^{'}$ of the extracted $p_{T}$ spectra of strange quarks. We see that the curvatures of strange quark spectra in both collisions are positive in the low $p_{T}$ range and turn to be negative at larger $p_{T}$, which means that the $\Omega/\phi$ ratio will firstly increase with $p_{T}$ and then decrease with $p_{T}$ and therefore exhibits an overall non-monotonic $p_{T}$ dependency. Comparing the curvature of strange quark spectrum in central Pb-Pb collisions at $\sqrt{s_{NN}}=$ 2.76 TeV with that in high-multiplicity $pp$ collisions at $\sqrt{s}=$ 13 TeV, we see that the former in the low $p_{T}$ range is obviously higher than the latter, which means $\Omega/\phi$ ratio in central Pb-Pb collisions will increase faster than that in $pp$ collisions.  In addition, we see that the $p_{T}$ position for zero curvature of strange quark spectrum in Pb-Pb collisions is larger than that in $pp$ collisions about 0.5 GeV/$c$, which means the increase tendency of $\Omega/\phi$ ratio in Pb-Pb collisions can extend to higher $p_{T}$.

\begin{figure}[h]
\centering
\includegraphics[width=0.9\linewidth]{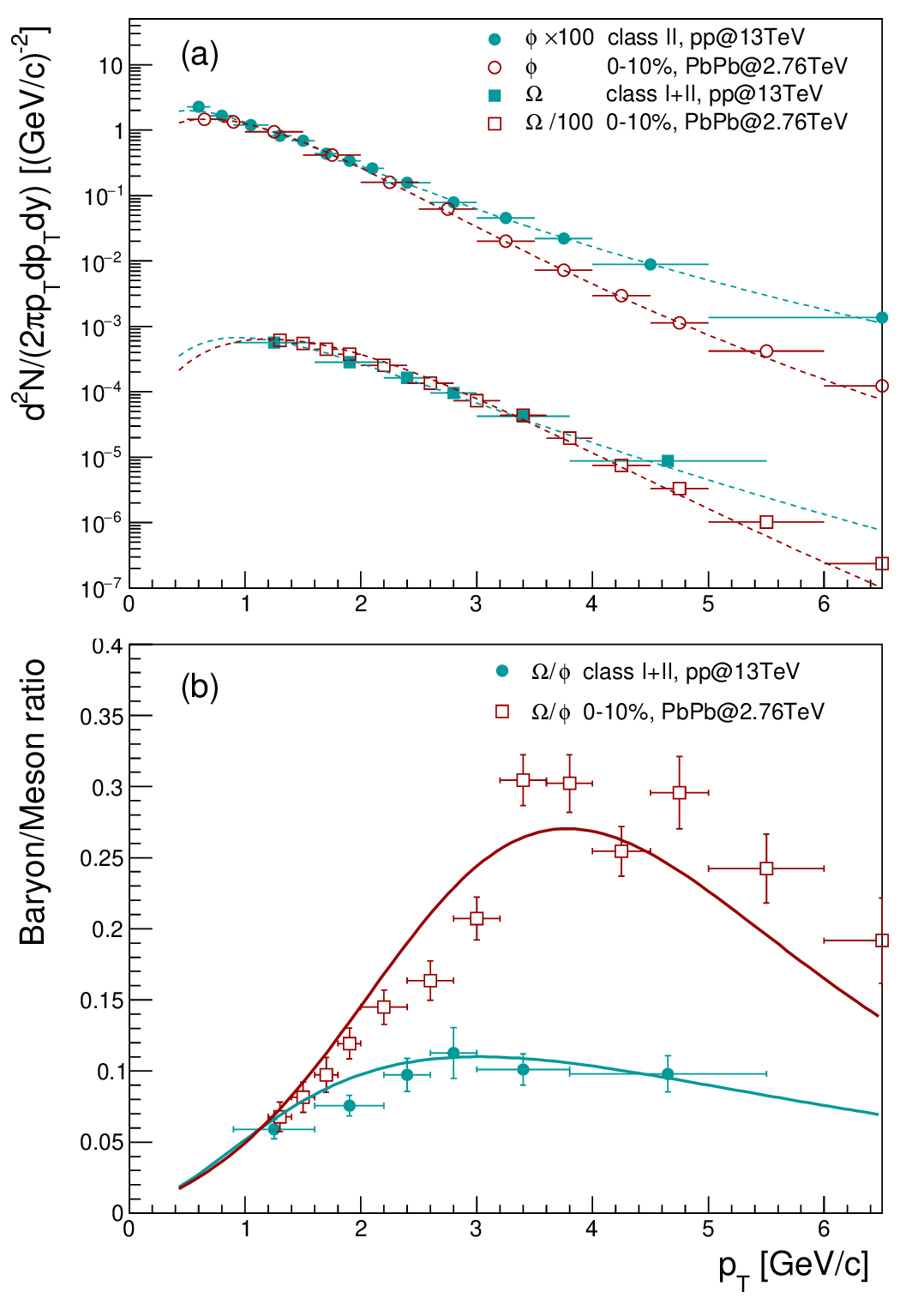}
    \caption{(a) $p_{T}$ spectra of $\Omega$ and $\phi$ in $pp$ collisions at $\sqrt{s}=13$ TeV and in Pb-Pb collisions at $\sqrt{s_{NN}}=$2.76 TeV and (b) the $\Omega/\phi$ ratio in the two collision systems.  Symbols are experimental data \citep{ALICE:2019etb,ALICE:2017ban,ALICE:2013xmt,ALICE:2019avo} and lines are results of EVC model. }\label{fig:Omg_phi_ratio_evc_r1}
\end{figure}

In Fig.~\ref{fig:Omg_phi_ratio_evc_r1}, we show the $p_{T}$ spectra of $\phi$ and $\Omega$ calculated by our EVC model and the resulting $\Omega/\phi$ ratio as the function of $p_{T}$. $p_{T}$ spectra of strange quarks are taken from fitting results in Fig.~\ref{fig: fs_curvature}(a).  Strangeness suppression factor $\lambda_{s}$ is taken to be 0.32 in $pp$ collisions and 0.43 in Pb-Pb collisions. Parameters $C_{\phi}=0.35$ and $a=4.90$ are taken in both collision systems. As expected, $\Omega/\phi$ ratio in $pp$ collisions relatively slowly increases with $p_{T}$ and reaches the peak value of about 0.1 at $p_{T}\approx3.0$ GeV/$c$ and then turns to decrease at larger $p_{T}$. $\Omega/\phi$ ratio in central Pb-Pb collisions, due to more concave property as shown in Fig.~\ref{fig: fs_curvature}(b), rapidly increases with $p_{T}$ in the low $p_{T}$ range and reaches the peak value of about 0.25 at $p_{T}\approx3.7$ GeV/$c$ and then turns to decrease at larger $p_{T}$. 

\begin{figure*}
\centering
\includegraphics[width=0.95\linewidth]{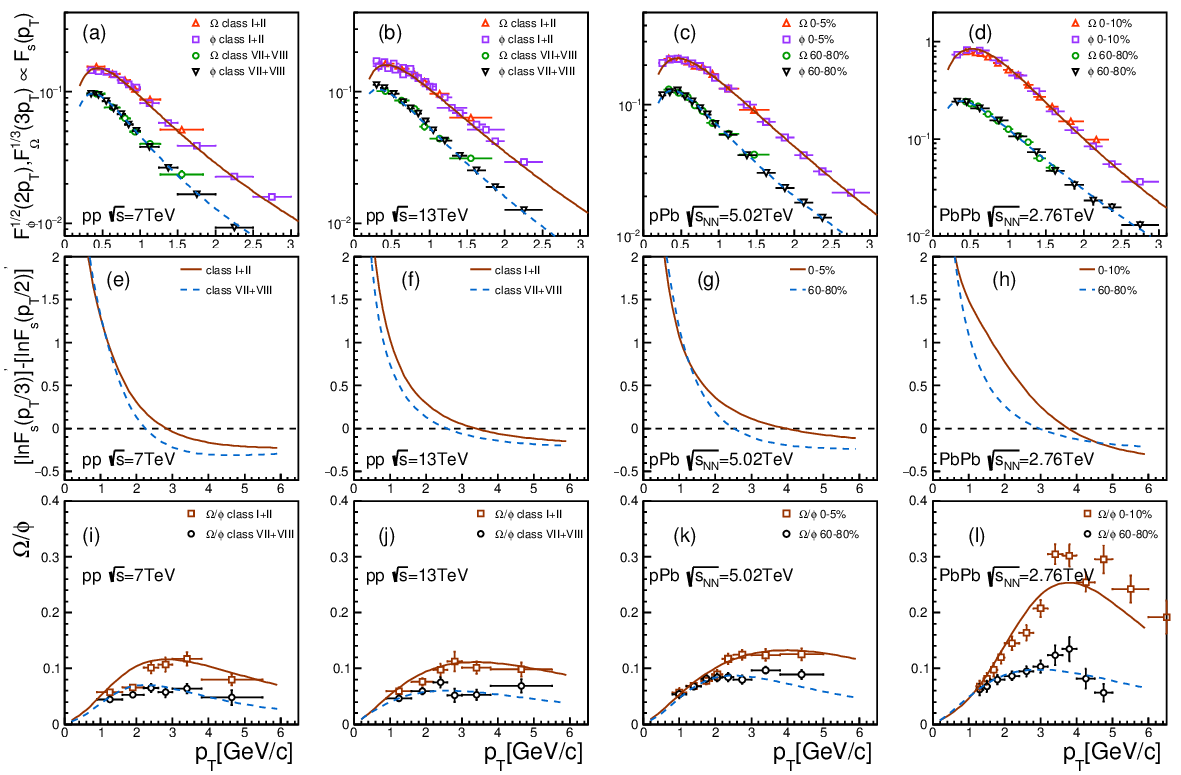}
    \caption{(a-d) the scaled $p_{T}$ spectra of $\Omega$ and $\phi$ in $pp$, $p$Pb and Pb-Pb collisions at LHC energies as a way to obtain $p_{T}$ distribution of strange quarks before hadronization;(e-h) the discrete curvature of $p_{T}$ distributions of strange quarks; (i-l) $\Omega/\phi$ ratios as the function of $p_{T}$. Symbols in panels (a-d) are experimental data \citep{ALICE:2020jsh,ALICE:2015mpp,ALICE:2019etb,ALICE:2019avo,ALICE:2018pal,ALICE:2017ban,ALICE:2016sak,ALICE:2013xmt} and lines are model results. }\label{fig:fs_dp2_Omg_phi_ratio_compare}
\end{figure*}

In Fig.~\ref{fig:fs_dp2_Omg_phi_ratio_compare}, we show a systematical analysis of $\Omega/\phi$ ratio as the function of $p_{T}$ in $pp$ collisions at $\sqrt{s}=$ 7, 13 TeV, $p$-Pb collisions at $\sqrt{s_{NN}}=$5.02 TeV and Pb-Pb collisions at $\sqrt{s_{NN}}=$2.76 TeV. A high multiplicity event class and a low multiplicity event class are selected in each collision system. Fig.~\ref{fig:fs_dp2_Omg_phi_ratio_compare}(a-d) show the scaled $p_{T}$ spectra of $\Omega$ and $\phi$ as a way to reveal $p_{T}$ distributions of strange quarks before hadronization. Data of $p_{T}$ spectra of $\Omega$ and $\phi$ are taken from \citep{ALICE:2013xmt,ALICE:2016sak,ALICE:2017ban,ALICE:2018pal,ALICE:2019avo,ALICE:2019etb,ALICE:2020jsh}.

Fig.~\ref{fig:fs_dp2_Omg_phi_ratio_compare}(e-h) show the discrete curvature $\left[\ln F_{s}(p_{T}/3)\right]^{'}-\left[\ln F_{s}(p_{T}/2)\right]^{'}$ of $p_{T}$ distributions of strange quarks which directly influences the $p_{T}$ dependence of $\Omega/\phi$ ratio shown in Fig.~\ref{fig:fs_dp2_Omg_phi_ratio_compare}(i-l).  In $\Omega/\phi$ ratio evaluation, parameters $C_{\phi}=0.35$ and $a=5.0$ are taken in all collision systems for simplicity. Strangeness factor $\lambda_{s}$ is taken to be 0.32 (0.29) in high (low) multiplicity events in $pp$ collisions at $\sqrt{s}=$ 7 and 13 TeV and is taken to be 0.35(0.29) in high (low) multiplicity events in $p$-Pb collisions at $\sqrt{s_{NN}}=$ 5.02 TeV and is taken to be 0.42 (0.30) in central (peripheral) Pb-Pb collisions $\sqrt{s_{NN}}=$ 2.76 TeV. 

Besides the production suppression of $\Omega$ relative to $\phi$ due to $\lambda_{s}$, the production of $\Omega$ in low multiplicity events will suffer more suppression because of the need of three strange quarks for a $\Omega$ formation. The average number of strange quarks at mid-rapidity in low multiplicity classes in $pp$ collisions is usually less than one \citep{Gou:2017foe,Zhang:2018vyr}, the formation of $\Omega$ in these event classes only happens in the part of events with three strange quarks and above. Therefore, we have to consider the event distribution of strange quark number to incorporate the event-by-event fluctuation of strange quark number in order to reproduce the yield of $\Omega$ in low multiplicity event classes \citep{Shao:2017eok,Gou:2017foe,Zhang:2018vyr}.  Here, we simplify this procedure by adding an extra suppression factor for $\Omega$ formation in low multiplicity event class since here we focus on the $p_{T}$ dependence of the $\Omega/\phi$ ratio. We take it to be 0.66 for multiplicity class VII+VIII in $pp$ collisions at $\sqrt{s}=$ 7 and 13 TeV and 0.85 for 60-80\% centrality in $p$-Pb collisions at $\sqrt{s_{NN}}=$ 5.02 TeV and Pb-Pb collisions $\sqrt{s_{NN}}=$ 2.76 TeV. 

We observe some features of $\Omega/\phi$ ratio in above four collision systems. Comparing $\Omega/\phi$ ratios in low and high multiplicity event classes in each collision system shown in Fig.~\ref{fig:fs_dp2_Omg_phi_ratio_compare}(i-l), we see a shift to higher $p_{T}$ for the peak position of $\Omega/\phi$ ratio in high multiplicity event class. This can be seen also from the curvature of strange quark distribution in Fig.~\ref{fig:fs_dp2_Omg_phi_ratio_compare}(e-h).  Comparing $\Omega/\phi$ ratios in high multiplicity event class in $pp$ collisions at $\sqrt{s}=$ 7 and 13 TeV and those in $p$Pb collisions at $\sqrt{s_{NN}}=$ 5.02 TeV, we see a weak increase of $\Omega/\phi$ ratio when the multiplicity of systems at mid-rapidity changes from 21.3 to 45 \citep{ALICE:2018pal,ALICE:2016sak}. However, we see a strong increase of $\Omega/\phi$ ratio in central Pb-Pb collisions, which can be also seen by higher curvature of strange quark distribution in low $p_{T}$ range in Fig.~\ref{fig:fs_dp2_Omg_phi_ratio_compare}(h). 

Considering $\phi$'s $\left\langle p_{T}\right\rangle =$ $1.440\pm0.023$, $1.456\pm0.015$ GeV/$c$ in highest multiplicity $pp$ collisions at $\sqrt{s}=$ 7 and 13 TeV \citep{ALICE:2019etb,ALICE:2018pal}, $\left\langle p_{T}\right\rangle =1.429\pm0.008$ GeV/$c$ in central (0-10\%) $p$-Pb collisions at $\sqrt{s_{NN}}=$ 5.02 TeV \citep{ALICE:2016sak} and $\left\langle p_{T}\right\rangle =$ $1.325\pm0.028$ GeV/$c$ in central Pb-Pb collisions at $\sqrt{s_{NN}}=2.76$ TeV \citep{ALICE:2017ban}, we see that $p_{T}$ spectra of $\phi$ (equivalently, strange quarks) in three small collision systems are globally flatter/broader than that in central Pb-Pb collisions but this does not induce the enhancement of $\Omega$ production at intermediate $p_{T}$. As we explained in this work, the curvature property of strange quark $p_{T}$ distribution is more relevant to the enhancement of $\Omega/\phi$ ratios at relatively intermediate $p_{T}$. This difference in curvature property of quark $p_{T}$ distribution should be related to the difference between the evolution property of parton system just before hadronization in small size and that in large size. 

The expansion evolution of bulk and hot parton medium produced in relativistic heavy-ion collisions is well modeled by relativistic hydrodynamics and strong collective flow is established in central collisions \citep{Kolb:2003dz,Gale:2013da}. The collective flow is also recently studied and observed in $pp$ and $p$-Pb collisions at LHC \citep{Zhao:2017rgg,Zhao:2020wcd,ALICE:2024vzv}. Here, we qualitatively demonstrate the influence of strong collective flow on the curvature of particle $p_{T}$ spectrum by the fit function in Eq.~(\ref{eq:fpt_levy_v3}) which is the Levy-Tsallis function boosted by a radial velocity $v$. The function has the asymptotic behavior of thermal distribution as $n\to\infty$. At finite $n$ the function also behaves like $\exp\left[-(m_{T}-p_{T}v-m)/\left(c\sqrt{1-v^{2}}\right)\right]$ at low $p_{T}$ which is the thermal distribution $\exp\left[-(m_{T}-m)/c\right]$ boosted with velocity $v$. Here $c$ can be roughly viewed as the temperature in the low $p_{T}$ range. The function behaves as power form $(1+p_{T}/a)^{-n}$ at large $p_{T}$, which is not influenced by $v$ in shape property. 

\begin{figure*}
\centering{}\includegraphics[width=0.85\linewidth]{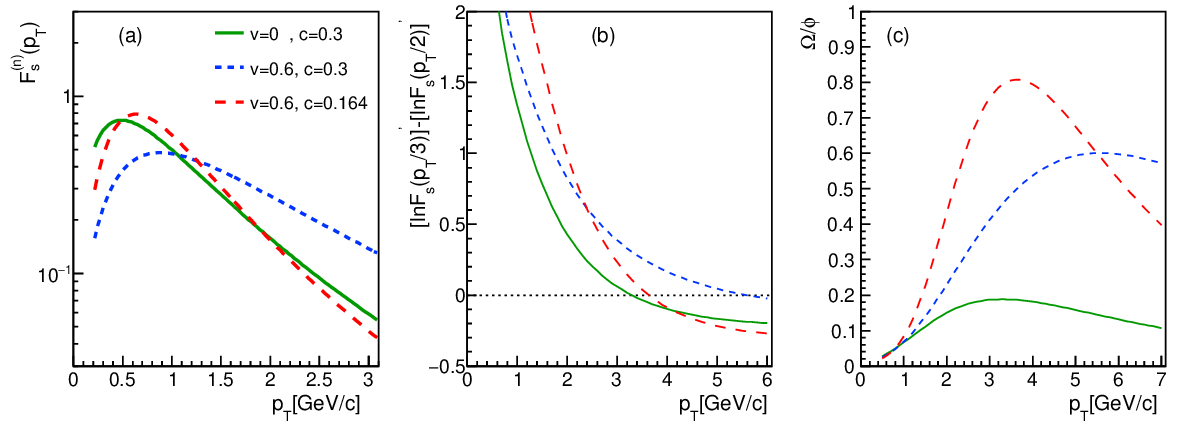}
    \caption{(a)The normalized $p_{T}$ spectrum of strange quarks under different
parameter values;(b) the curvature of the spectrum; (c) the schematic
behavior of $\Omega/\phi$ ratio with the setting of the same magnitude
at low $p_{T}$. }\label{fig:levy_dp2_Omg_phi_ratio_dem}
\end{figure*}

In Fig.~\ref{fig:levy_dp2_Omg_phi_ratio_dem}(a), we show examples of the function in Eq.~(\ref{eq:fpt_levy_v3}) for three different parameter groups. We select $v=0$, $c=0.3$ as the base line, which has the meaning of static source with relatively high temperature (plus the power form tail from jet physics). We select $v=0.6$, $c=0.3$ as the second example of high temperature source with a strong collective flow and select $v=0.6$, $c=0.164$ as the third example of low temperature source with a strong collective flow. The third example is to capture the expansion and cooling property of hot medium evolution and parameters are selected to meet that its $\left\langle p_{T}\right\rangle $ is the same as that in the base line. The corresponding curvatures of strange quark $p_{T}$ spectrum for above different parameter groups are shown in Fig.~\ref{fig:levy_dp2_Omg_phi_ratio_dem}(b). Fig.~\ref{fig:levy_dp2_Omg_phi_ratio_dem}(c) shows the schematic behavior of $\Omega/\phi$ ratios in three cases under the setting of the same magnitude at low $p_{T}$ in order to only demonstrate the influence of the curvature property of strange quark $p_{T}$ spectrum. We see that the $\Omega/\phi$ ratio in the base line case, the solid line in Fig.~\ref{fig:levy_dp2_Omg_phi_ratio_dem}(c), increases slowly at low $p_{T}$ and turns to decrease with $p_{T}$ as $p_{T}\gtrsim3.0$ GeV/$c$. The $\Omega/\phi$ ratio in the case of high temperature and large collective flow, the dashed line, increases rapidly with $p_{T}$ and turns to decrease with $p_{T}$ at larger $p_{T}\gtrsim5.5$ GeV/$c$. The $\Omega/\phi$ ratio in the case of low temperature and large collective flow, the long dashed line, increases most rapidly with $p_{T}$ in the low $p_{T}$ range and can reach a quite high value at intermediate $p_{T}$ and finally turns to decrease with $p_{T}$. 

\section{Summary and discussion}\label{sec:summary}

In this paper, we have applied the constituent quark equal-velocity combination model to study the $p_{T}$ dependence of $\Omega/\phi$ ratio in $pp$, $p$-Pb and Pb-Pb collisions at LHC energies. In this model, $p_{T}$ distributions of the produced $\Omega$ and $\phi$ are simple product of $p_{T}$ spectrum of strange quarks just before hadronization. Therefore, we can derive the analytical result of $\Omega/\phi$ ratio as the function of $p_{T}$ and find out the key influence ingredient of the $p_{T}$ dependence of $\Omega/\phi$ ratio, i.e., the discrete curvature of strange quark $p_{T}$ spectrum. According to the quark number scaling property, we extracted the $p_{T}$ spectra of strange quarks from experimental data of $\phi$ and studied the curvature property of $p_{T}$ spectra of strange quarks in high and low multiplicity events in $pp$ collisions at $\sqrt{s}=$ 7, 13 TeV, $p$-Pb collisions at $\sqrt{s_{NN}}$= 5.02 TeV and Pb-Pb collisions at $\sqrt{s_{NN}}$= 2.76 TeV. Finally, we applied these curvature properties of $p_{T}$ spectra of strange quarks to successfully explain the observed $p_{T}$ dependence of $\Omega/\phi$ ratio in these collisions. 

One implication of this work is that the geometry property of the $p_{T}$ distribution of quarks is a useful information to understand the hadron production in high energy collisions.In this paper, we build a simple connection between the discrete curvature of strange quark $p_{T}$ distribution just before hadronization and the $p_{T}$ dependence of $\Omega/\phi$ ratio. By comparing the curvature of strange quark distributions in $pp$, $p$-Pb and Pb-Pb, we find a relatively large change for the curvature of quark distribution when the collision system changes from $pp$, $p$-Pb to Pb-Pb collisions.  This large change can be qualitatively understood as the consequence of strong collective flow such as that formed in central AA collisions. 

Similar curvature property analysis can be also applied to $p_{T}$ distribution of hadrons. In our latest work \citep{Song:2025oay}, we observed a systematical difference between curvatures of baryon distributions and those of mesons which can be explained via quark combination mechanism at hadronization. Also, correlations between the characteristic points of the curvature of hadronic $p_{T}$ distributions and the measured $\left\langle p_{T}\right\rangle $ of hadrons in $pp$, $p$-A and AA collisions also suggest an obvious change relating to the size (charged-particle multiplicity) of collision system. This property is also in close connection to the influence of collective flow. In summary, we argue that some quantification of the local geometry property of $p_{T}$ distribution of particles is a new and effective way of understanding particle production in high energy collisions.  

\begin{acknowledgments} 
    This work is supported by Shandong Provincial Natural Science Foundation (Grants No. ZR2025MS01) and by the National Natural Science Foundation of China under Grants No. 12375074. 
\end{acknowledgments}

\bibliographystyle{apsrev4-1}
\bibliography{ref}

@article{ALICE:2010mlf,
    author = "Aamodt, Kenneth and others",
    collaboration = "ALICE",
    title = "{Centrality dependence of the charged-particle multiplicity density at mid-rapidity in Pb-Pb collisions at $\sqrt{s_{NN}}=2.76$ TeV}",
    eprint = "1012.1657",
    archivePrefix = "arXiv",
    primaryClass = "nucl-ex",
    reportNumber = "CERN-PH-EP-2010-071",
    doi = "10.1103/PhysRevLett.106.032301",
    journal = "Phys. Rev. Lett.",
    volume = "106",
    pages = "032301",
    year = "2011"
}

@article{ALICE:2019avo,
    author = "Acharya, Shreyasi and others",
    collaboration = "ALICE",
    title = "{Multiplicity dependence of (multi-)strange hadron production in proton-proton collisions at $\sqrt{s}$ = 13 TeV}",
    eprint = "1908.01861",
    archivePrefix = "arXiv",
    primaryClass = "nucl-ex",
    reportNumber = "CERN-EP-2019-168",
    doi = "10.1140/epjc/s10052-020-7673-8",
    journal = "Eur. Phys. J. C",
    volume = "80",
    number = "2",
    pages = "167",
    year = "2020"
}

@article{ALICE:2019etb,
    author = "Acharya, Shreyasi and others",
    collaboration = "ALICE",
    title = "{Multiplicity dependence of K*(892)0 and {\ensuremath{\phi}}(1020) production in pp collisions at s=13 TeV}",
    eprint = "1910.14397",
    archivePrefix = "arXiv",
    primaryClass = "nucl-ex",
    reportNumber = "CERN-EP-2019-245",
    doi = "10.1016/j.physletb.2020.135501",
    journal = "Phys. Lett. B",
    volume = "807",
    pages = "135501",
    year = "2020"
}

@article{ALICE:2017ban,
    author = "Adam, Jaroslav and others",
    collaboration = "ALICE",
    title = "{K$^{*}(892)^{0}$ and $\phi(1020)$ meson production at high transverse momentum in pp and Pb-Pb collisions at $\sqrt{s_\mathrm{NN}}$ = 2.76 TeV}",
    eprint = "1702.00555",
    archivePrefix = "arXiv",
    primaryClass = "nucl-ex",
    reportNumber = "CERN-EP-2017-010",
    doi = "10.1103/PhysRevC.95.064606",
    journal = "Phys. Rev. C",
    volume = "95",
    number = "6",
    pages = "064606",
    year = "2017"
}

@article{ALICE:2013xmt,
    author = "Abelev, Betty Bezverkhny and others",
    collaboration = "ALICE",
    title = "{Multi-strange baryon production at mid-rapidity in Pb-Pb collisions at $\sqrt{s_{NN}}$ = 2.76 TeV}",
    eprint = "1307.5543",
    archivePrefix = "arXiv",
    primaryClass = "nucl-ex",
    reportNumber = "CERN-PH-EP-2013-134",
    doi = "10.1016/j.physletb.2014.05.052",
    journal = "Phys. Lett. B",
    volume = "728",
    pages = "216--227",
    year = "2014",
    note = "[Erratum: Phys.Lett.B 734, 409--410 (2014)]"
}

@article{ALICE:2020jsh,
    author = "Acharya, Shreyasi and others",
    collaboration = "ALICE",
    title = "{Production of light-flavor hadrons in pp collisions at $\sqrt{s}~=~7\text { and }\sqrt{s} = 13 \, \text { TeV} $}",
    eprint = "2005.11120",
    archivePrefix = "arXiv",
    primaryClass = "nucl-ex",
    reportNumber = "CERN-EP-2020-059",
    doi = "10.1140/epjc/s10052-020-08690-5",
    journal = "Eur. Phys. J. C",
    volume = "81",
    number = "3",
    pages = "256",
    year = "2021"
}

@article{Tsallis:1987eu,
    author = "Tsallis, Constantino",
    title = "{Possible Generalization of Boltzmann-Gibbs Statistics}",
    reportNumber = "CBPF-NF-062-87",
    doi = "10.1007/BF01016429",
    journal = "J. Statist. Phys.",
    volume = "52",
    pages = "479--487",
    year = "1988"
}

@article{Song:2017gcz,
      author         = "Song, Jun and Gou, Xing-rui and Shao, Feng-lan and Liang,
                        Zuo-Tang",
      title          = "{Quark number scaling of hadronic $p_T$ spectra and
                        constituent quark degree of freedom in $p$-Pb collisions
                        at $\sqrt{s_{NN}}=5.02$ TeV}",
      journal        = "Phys. Lett.",
      volume         = "B774",
      year           = "2017",
      pages          = "516-521",
      doi            = "10.1016/j.physletb.2017.10.012",
      eprint         = "1707.03949",
      archivePrefix  = "arXiv",
      primaryClass   = "hep-ph",
      SLACcitation   = "%%CITATION = ARXIV:1707.03949;%%"
}

@article{Gou:2017foe,
      author         = "Gou, Xing-rui and Shao, Feng-lan and Wang, Rui-qin and
                        Li, Hai-hong and Song, Jun",
      title          = "{New insights into hadron production mechanism from
                        $p_{T}$ spectra in $pp$ collisions at $\sqrt{s}=7$ TeV}",
      journal        = "Phys. Rev.",
      volume         = "D96",
      year           = "2017",
      number         = "9",
      pages          = "094010",
      doi            = "10.1103/PhysRevD.96.094010",
      eprint         = "1707.06906",
      archivePrefix  = "arXiv",
      primaryClass   = "hep-ph",
      SLACcitation   = "%%CITATION = ARXIV:1707.06906;%%"
}

@article{Song:2019sez,
    author = "Song, Jun and Shao, Feng-lan and Liang, Zuo-tang",
    title = "{Quark number scaling of $p_{T}$ spectra for $\Omega$ and $\phi$ in relativistic heavy-ion collisions}",
    eprint = "1911.01152",
    archivePrefix = "arXiv",
    primaryClass = "nucl-th",
    doi = "10.1103/PhysRevC.102.014911",
    journal = "Phys. Rev. C",
    volume = "102",
    number = "1",
    pages = "014911",
    year = "2020"
}

@article{Song:2020kak,
    author = "Song, Jun and Wang, Xiao-Feng and Li, Hai-Hong and Wang, Rui-Qin and Shao, Feng-Lan",
    title = "{Strange hadron production in a quark combination model in Au+Au collisions at energies available at the BNL Relativistic Heavy Ion Collider}",
    eprint = "2007.14588",
    archivePrefix = "arXiv",
    primaryClass = "nucl-th",
    doi = "10.1103/PhysRevC.103.034907",
    journal = "Phys. Rev. C",
    volume = "103",
    number = "3",
    pages = "034907",
    year = "2021"
}

@article{Shao:2017eok,
    author = "Shao, Feng-lan and Wang, Guo-jing and Wang, Rui-qin and Li, Hai-hong and Song, Jun",
    archivePrefix = "arXiv",
    doi = "10.1103/PhysRevC.95.064911",
    eprint = "1703.05862",
    journal = "Phys.\ Rev.\ C",
    number = "6",
    pages = "064911",
    primaryClass = "hep-ph",
    title = "{Yield ratios of identified hadrons in p+p, p+Pb , and Pb+Pb collisions at energies available at the CERN Large Hadron Collider}",
    volume = "95",
    year = "2017"
}

@article{Song:2013isa,
    author = "Song, Jun and Shao, Feng-lan",
    archivePrefix = "arXiv",
    doi = "10.1103/PhysRevC.88.027901",
    eprint = "1303.1231",
    journal = "Phys.\ Rev.\ C",
    pages = "027901",
    primaryClass = "nucl-th",
    title = "{Baryon-antibaryon production asymmetry in relativistic heavy ion collisions}",
    volume = "88",
    year = "2013"
}

@article{Zhang:2018vyr,
	author         = "Zhang, Jian-wei and Li, Hai-hong and Shao, Feng-lan and
		Song, Jun",
	title          = "{Constituent quark number scaling from strange hadron spectra in $pp$
		collisions at $\sqrt{s}=$ 13 TeV}",
	journal = "Chin. Phys.",
	volume = "C44",
	number = "1",
	pages = "014101",
	year = "2020",
	doi = "10.1088/1674-1137/44/1/014101",
	eprint         = "1811.00975",
	archivePrefix  = "arXiv",
	primaryClass   = "hep-ph",
	SLACcitation   = "%%CITATION = ARXIV:1811.00975;%%"
}

@article{ALICE:2018pal,
    author = "Acharya, Shreyasi and others",
    collaboration = "ALICE",
    title = "{Multiplicity dependence of light-flavor hadron production in pp collisions at $\sqrt{s}$ = 7 TeV}",
    eprint = "1807.11321",
    archivePrefix = "arXiv",
    primaryClass = "nucl-ex",
    reportNumber = "CERN-EP-2018-209",
    doi = "10.1103/PhysRevC.99.024906",
    journal = "Phys. Rev. C",
    volume = "99",
    number = "2",
    pages = "024906",
    year = "2019"
}

@article{ALICE:2016sak,
    author = "Adam, Jaroslav and others",
    collaboration = "ALICE",
    title = "{Production of K$^{*}$ (892)$^{0}$ and $\phi $ (1020) in p{\textendash}Pb collisions at $\sqrt{s_{{\text {NN}}}}$ = 5.02 TeV}",
    eprint = "1601.07868",
    archivePrefix = "arXiv",
    primaryClass = "nucl-ex",
    reportNumber = "CERN-PH-EP-2015-326",
    doi = "10.1140/epjc/s10052-016-4088-7",
    journal = "Eur. Phys. J. C",
    volume = "76",
    number = "5",
    pages = "245",
    year = "2016"
}

@article{Zhao:2017rgg,
    author = "Zhao, Wenbin and Zhou, You and Xu, Haojie and Deng, Weitian and Song, Huichao",
    title = "{Hydrodynamic collectivity in proton\textendash{}proton collisions at 13 TeV}",
    eprint = "1801.00271",
    archivePrefix = "arXiv",
    primaryClass = "nucl-th",
    doi = "10.1016/j.physletb.2018.03.022",
    journal = "Phys. Lett. B",
    volume = "780",
    pages = "495--500",
    year = "2018"
}

@article{ALICE:2024vzv,
    author = "Acharya, Shreyasi and others",
    collaboration = "ALICE",
    title = "{Observation of partonic flow in proton{\textemdash}proton and proton{\textemdash}nucleus collisions}",
    eprint = "2411.09323",
    archivePrefix = "arXiv",
    primaryClass = "nucl-ex",
    reportNumber = "CERN-EP-2024-299",
    doi = "10.1038/s41467-025-67795-1",
    journal = "Nature Commun.",
    volume = "17",
    number = "1",
    pages = "2585",
    year = "2026"
}

@article{Zhao:2020wcd,
    author = "Zhao, Wenbin and Ko, Che Ming and Liu, Yu-Xin and Qin, Guang-You and Song, Huichao",
    title = "{Probing the Partonic Degrees of Freedom in High-Multiplicity $p-Pb$ collisions at $\sqrt {s_{NN}}$ = 5.02  TeV}",
    eprint = "1911.00826",
    archivePrefix = "arXiv",
    primaryClass = "nucl-th",
    doi = "10.1103/PhysRevLett.125.072301",
    journal = "Phys. Rev. Lett.",
    volume = "125",
    number = "7",
    pages = "072301",
    year = "2020"
}

@article{Gale:2013da,
    author = "Gale, Charles and Jeon, Sangyong and Schenke, Bjoern",
    title = "{Hydrodynamic Modeling of Heavy-Ion Collisions}",
    eprint = "1301.5893",
    archivePrefix = "arXiv",
    primaryClass = "nucl-th",
    doi = "10.1142/S0217751X13400113",
    journal = "Int. J. Mod. Phys. A",
    volume = "28",
    pages = "1340011",
    year = "2013"
}

@inbook{Kolb:2003dz,
    author = "Kolb, Peter F. and Heinz, Ulrich W.",
    editor = "Hwa, Rudolph C. and Wang, Xin-Nian",
    title = "{Hydrodynamic description of ultrarelativistic heavy ion collisions}",
    eprint = "nucl-th/0305084",
    archivePrefix = "arXiv",
	booktitle = {Quark–Gluon Plasma 3},
	publisher = {WORLD SCIENTIFIC},
	chapter = {},
    pages = "634--714",
    month = "5",
	doi = {10.1142/9789812795533_0010},
    year = "2003"
}

@article{ALICE:2015mpp,
    author = "Adam, Jaroslav and others",
    collaboration = "ALICE",
    title = "{Multi-strange baryon production in p-Pb collisions at $\sqrt{s_\mathbf{NN}}=5.02$ TeV}",
    eprint = "1512.07227",
    archivePrefix = "arXiv",
    primaryClass = "nucl-ex",
    reportNumber = "CERN-PH-EP-2015-327",
    doi = "10.1016/j.physletb.2016.05.027",
    journal = "Phys. Lett. B",
    volume = "758",
    pages = "389--401",
    year = "2016"
}

@article{Li:2021nhq,
    author = "Li, Hai-hong and Shao, Feng-lan and Song, Jun",
    title = "{Production of light-flavor and single-charmed hadrons in pp collisions at TeV in an equal-velocity quark combination model}",
    eprint = "2103.14900",
    archivePrefix = "arXiv",
    primaryClass = "hep-ph",
    doi = "10.1088/1674-1137/ac1ef9",
    journal = "Chin. Phys. C",
    volume = "45",
    number = "11",
    pages = "113105",
    year = "2021"
}

@article{Chen:2006vc,
    author = "Chen, Lie-Wen and Ko, Che Ming",
    title = "{phi and omega production from relativistic heavy ion collisions in a dynamical quark coalescence model}",
    eprint = "nucl-th/0602025",
    archivePrefix = "arXiv",
    doi = "10.1103/PhysRevC.73.044903",
    journal = "Phys. Rev. C",
    volume = "73",
    pages = "044903",
    year = "2006"
}

@article{STAR:2008bgi,
    author = "Abelev, B. I. and others",
    collaboration = "STAR",
    title = "{Measurements of phi meson production in relativistic heavy-ion collisions at RHIC}",
    eprint = "0809.4737",
    archivePrefix = "arXiv",
    primaryClass = "nucl-ex",
    doi = "10.1103/PhysRevC.79.064903",
    journal = "Phys. Rev. C",
    volume = "79",
    pages = "064903",
    year = "2009"
}

@article{Pu:2018eei,
    author = "Pu, Jie and Sun, Kai-Jia and Chen, Lie-Wen",
    title = "{Extracting strange quark freeze-out information in Pb+Pb collisions at $\sqrt{s_{NN}}$=2.76 TeV from $\phi$ and $\Omega$ production}",
    eprint = "1808.04053",
    archivePrefix = "arXiv",
    primaryClass = "nucl-th",
    doi = "10.1103/PhysRevC.98.064905",
    journal = "Phys. Rev. C",
    volume = "98",
    number = "6",
    pages = "064905",
    year = "2018"
}

@article{Jin:2018lbk,
    author = "Jin, Xiao-Hai and Chen, Jin-Hui and Ma, Yu-Gang and Zhang, Song and Zhang, Chun-Jian and Zhong, Chen",
    title = "{{\ensuremath{\Omega}} and {\ensuremath{\phi}} production in Au + Au collisions at $\sqrt{s_{_{\mathrm{NN}}}}=11.5$   and...}",
    doi = "10.1007/s41365-018-0393-1",
    journal = "Nucl. Sci. Tech.",
    volume = "29",
    number = "4",
    pages = "54",
    year = "2018"
}

@article{Hwa:2006vb,
    author = "Hwa, Rudolph C. and Yang, C. B.",
    title = "{Production of strange particles at intermediate pT in central Au+Au collisions at high energies}",
    eprint = "nucl-th/0602024",
    archivePrefix = "arXiv",
    doi = "10.1103/PhysRevC.75.054904",
    journal = "Phys. Rev. C",
    volume = "75",
    pages = "054904",
    year = "2007"
}

@proceedings{Proceedings:2007ctk,
    author = "Abreu, S. and others",
    editor = "Armesto, N. and Borghini, N. and Jeon, S. and Wiedemann, U. A.",
    title = "{Proceedings, Workshop on Heavy Ion Collisions at the LHC: Last Call for Predictions}: {Geneva, Switzerland, May 14 - June 8, 2007}",
    eprint = "0711.0974",
    archivePrefix = "arXiv",
    primaryClass = "hep-ph",
    doi = "10.1088/0954-3899/35/5/054001",
    volume = "35",
    pages = "054001",
    year = "2008"
}

@article{ALICE:2014jbq,
    author = "Abelev, Betty Bezverkhny and others",
    collaboration = "ALICE",
    title = "{$K^*(892)^0$ and $ϕ(1020)$ production in Pb-Pb collisions at $\sqrt{s{NN}}$ = 2.76 TeV}",
    eprint = "1404.0495",
    archivePrefix = "arXiv",
    primaryClass = "nucl-ex",
    reportNumber = "CERN-PH-EP-2014-060",
    doi = "10.1103/PhysRevC.91.024609",
    journal = "Phys. Rev. C",
    volume = "91",
    pages = "024609",
    year = "2015"
}

@article{Shor:1984ui,
    author = "Shor, A.",
    title = "{Phi meson production as a probe of the quark gluon plasma}",
    doi = "10.1103/PhysRevLett.54.1122",
    journal = "Phys. Rev. Lett.",
    volume = "54",
    pages = "1122--1125",
    year = "1985"
}

@article{vanHecke:1998yu,
    author = "van Hecke, H. and Sorge, H. and Xu, N.",
    title = "{Evidence of early multistrange hadron freezeout in high-energy nuclear collisions}",
    eprint = "nucl-th/9804035",
    archivePrefix = "arXiv",
    doi = "10.1103/PhysRevLett.81.5764",
    journal = "Phys. Rev. Lett.",
    volume = "81",
    pages = "5764--5767",
    year = "1998"
}

@article{PHENIX:2001vgc,
    author = "Adcox, K. and others",
    collaboration = "PHENIX",
    title = "{Centrality dependence of pi+ / pi-, K+ / K-, p and anti-p production from s(NN)**(1/2) = 13-=GeV Au+Au collisions at RHIC}",
    eprint = "nucl-ex/0112006",
    archivePrefix = "arXiv",
    doi = "10.1103/PhysRevLett.88.242301",
    journal = "Phys. Rev. Lett.",
    volume = "88",
    pages = "242301",
    year = "2002"
}

@article{STAR:2006uve,
    author = "Abelev, B. I. and others",
    collaboration = "STAR",
    title = "{Identified baryon and meson distributions at large transverse momenta from Au+Au collisions at s(NN)**(1/2) = 200-GeV}",
    eprint = "nucl-ex/0606003",
    archivePrefix = "arXiv",
    doi = "10.1103/PhysRevLett.97.152301",
    journal = "Phys. Rev. Lett.",
    volume = "97",
    pages = "152301",
    year = "2006"
}

@article{Greco:2003xt,
    author = "Greco, V. and Ko, C. M. and Levai, P.",
    title = "{Parton coalescence and anti-proton / pion anomaly at RHIC}",
    eprint = "nucl-th/0301093",
    archivePrefix = "arXiv",
    doi = "10.1103/PhysRevLett.90.202302",
    journal = "Phys. Rev. Lett.",
    volume = "90",
    pages = "202302",
    year = "2003"
}

@article{Hwa:2002tu,
    author = "Hwa, Rudolph C. and Yang, C. B.",
    title = "{Scaling behavior at high p(T) and the p / pi ratio}",
    eprint = "nucl-th/0211010",
    archivePrefix = "arXiv",
    reportNumber = "OITS-720",
    doi = "10.1103/PhysRevC.67.034902",
    journal = "Phys. Rev. C",
    volume = "67",
    pages = "034902",
    year = "2003"
}

@article{ALICE:2017thy,
    author = "Acharya, Shreyasi and others",
    collaboration = "ALICE",
    title = "{$\Lambda_{\rm c}^+$ production in pp collisions at $\sqrt{s} = 7$ TeV and in p-Pb collisions at $\sqrt{s_{\rm NN}} = 5.02$ TeV}",
    eprint = "1712.09581",
    archivePrefix = "arXiv",
    primaryClass = "nucl-ex",
    reportNumber = "CERN-EP-2017-339",
    doi = "10.1007/JHEP04(2018)108",
    journal = "JHEP",
    volume = "04",
    pages = "108",
    year = "2018"
}

@article{Song:2023nzu,
    author = "Song, Jun and Li, Hai-hong and Shao, Feng-lan",
    title = "{Transverse momentum and multiplicity dependence of $\varLambda _{c}^{+}/D^{0}$ ratio in pp collisions at $\sqrt{s}=13$~TeV}",
    eprint = "2304.00434",
    archivePrefix = "arXiv",
    primaryClass = "hep-ph",
    doi = "10.1140/epjc/s10052-023-12007-7",
    journal = "Eur. Phys. J. C",
    volume = "83",
    number = "9",
    pages = "852",
    year = "2023"
}

@article{Song:2025oay,
    author = "Song, Jun and Li, Hai-hong and Shao, Feng-lan",
    title = "{Domain of soft hadrons in transverse momentum space in high energy collisions}",
    eprint = "2512.15178",
    archivePrefix = "arXiv",
    primaryClass = "hep-ph",
    doi = "10.1103/9jjq-4pqs",
    journal = "Phys. Rev. D",
    volume = "112",
    number = "11",
    pages = "114025",
    year = "2025"
}

@article{ALICE:2022exq,
    author = "Acharya, Shreyasi and others",
    collaboration = "ALICE",
    title = "{First measurement of {\ensuremath{\Lambda}}c+ production down to pT=0 in pp and p-Pb collisions at sNN=5.02 TeV}",
    eprint = "2211.14032",
    archivePrefix = "arXiv",
    primaryClass = "nucl-ex",
    reportNumber = "CERN-EP-2022-261",
    doi = "10.1103/PhysRevC.107.064901",
    journal = "Phys. Rev. C",
    volume = "107",
    number = "6",
    pages = "064901",
    year = "2023"
}

@article{LHCb:2022ddg,
    author = "Aaij, R. and others",
    collaboration = "LHCb",
    title = "{Measurement of the $ {\Lambda}_c^{+} $ to D$^{0}$ production ratio in peripheral PbPb collisions at $ \sqrt{s_{\textrm{NN}}} $ = 5.02 TeV}",
    eprint = "2210.06939",
    archivePrefix = "arXiv",
    primaryClass = "hep-ex",
    reportNumber = "LHCb-PAPER-2021-046, CERN-EP-2022-148",
    doi = "10.1007/JHEP06(2023)132",
    journal = "JHEP",
    volume = "06",
    pages = "132",
    year = "2023",
    note = "[Erratum: JHEP 05, 021 (2024)]"
}

@article{CMS:2023frs,
    author = "Tumasyan, Armen and others",
    collaboration = "CMS",
    title = "{Study of charm hadronization with prompt $ {\Lambda}_{\textrm{c}}^{+} $ baryons in proton-proton and lead-lead collisions at $ \sqrt{s_{\textrm{NN}}} $ = 5.02 TeV}",
    eprint = "2307.11186",
    archivePrefix = "arXiv",
    primaryClass = "nucl-ex",
    reportNumber = "CMS-HIN-21-004, CERN-EP-2023-085",
    doi = "10.1007/JHEP01(2024)128",
    journal = "JHEP",
    volume = "01",
    pages = "128",
    year = "2024"
}

@article{LHCb:2018weo,
    author = "Aaij, Roel and others",
    collaboration = "LHCb",
    title = "{Prompt $\Lambda^+_c$ production in $p\mathrm{Pb}$ collisions at $\sqrt{s_{NN}} = 5.02$ TeV}",
    eprint = "1809.01404",
    archivePrefix = "arXiv",
    primaryClass = "hep-ex",
    reportNumber = "LHCb-PAPER-2018-021, CERN-EP-2018-193",
    doi = "10.1007/JHEP02(2019)102",
    journal = "JHEP",
    volume = "02",
    pages = "102",
    year = "2019"
}

@article{Altmann:2024kwx,
    author = "Altmann, J. and Dubla, A. and Greco, V. and Rossi, A. and Skands, P.",
    title = "{Towards the understanding of heavy quarks hadronization: from leptonic to heavy-ion collisions}",
    eprint = "2405.19137",
    archivePrefix = "arXiv",
    primaryClass = "hep-ph",
    doi = "10.1140/epjc/s10052-024-13641-5",
    journal = "Eur. Phys. J. C",
    volume = "85",
    number = "1",
    pages = "16",
    year = "2025"
}

@article{Zhao:2023nrz,
    author = "Zhao, Jiaxing and others",
    title = "{Hadronization of heavy quarks}",
    eprint = "2311.10621",
    archivePrefix = "arXiv",
    primaryClass = "hep-ph",
    doi = "10.1103/PhysRevC.109.054912",
    journal = "Phys. Rev. C",
    volume = "109",
    number = "5",
    pages = "054912",
    year = "2024"
}

@article{Minissale:2020bif,
    author = "Minissale, Vincenzo and Plumari, Salvatore and Greco, Vincenzo",
    title = "{Charm hadrons in pp collisions at LHC energy within a coalescence plus fragmentation approach}",
    eprint = "2012.12001",
    archivePrefix = "arXiv",
    primaryClass = "hep-ph",
    doi = "10.1016/j.physletb.2021.136622",
    journal = "Phys. Lett. B",
    volume = "821",
    pages = "136622",
    year = "2021"
}

@article{STAR:2006nmo,
    author = "Abelev, B. I. and others",
    collaboration = "STAR",
    title = "{Strange particle production in p+p collisions at s**(1/2) = 200-GeV}",
    eprint = "nucl-ex/0607033",
    archivePrefix = "arXiv",
    doi = "10.1103/PhysRevC.75.064901",
    journal = "Phys. Rev. C",
    volume = "75",
    pages = "064901",
    year = "2007"
}

@article{STAR:2006pcq,
    author = "Adams, John and others",
    collaboration = "STAR, STAR RICH",
    title = "{Measurements of identified particles at intermediate transverse momentum in the STAR experiment from Au + Au collisions at $\sqrt{s_{NN}}=200$ GeV}",
    eprint = "nucl-ex/0601042",
    archivePrefix = "arXiv",
    month = "1",
    year = "2006"
}

@article{STAR:2019bjj,
    author = "Adam, Jaroslav and others",
    collaboration = "STAR",
    title = "{Strange hadron production in Au+Au collisions at $\sqrt{s_{_{\mathrm{NN}}}}$ = 7.7, 11.5, 19.6, 27, and 39 GeV}",
    eprint = "1906.03732",
    archivePrefix = "arXiv",
    primaryClass = "nucl-ex",
    doi = "10.1103/PhysRevC.102.034909",
    journal = "Phys. Rev. C",
    volume = "102",
    number = "3",
    pages = "034909",
    year = "2020"
}

@article{ALICE:2013wgn,
    author = "Abelev, Betty Bezverkhny and others",
    collaboration = "ALICE",
    title = "{Multiplicity Dependence of Pion, Kaon, Proton and Lambda Production in p-Pb Collisions at $\sqrt{s_{NN}}$ = 5.02 TeV}",
    eprint = "1307.6796",
    archivePrefix = "arXiv",
    primaryClass = "nucl-ex",
    reportNumber = "CERN-PH-EP-2013-135",
    doi = "10.1016/j.physletb.2013.11.020",
    journal = "Phys. Lett. B",
    volume = "728",
    pages = "25--38",
    year = "2014"
}

@article{ALICE:2014juv,
    author = "Abelev, Betty Bezverkhny and others",
    collaboration = "ALICE",
    title = "{Production of charged pions, kaons and protons at large transverse momenta in pp and Pb{\textendash}Pb collisions at $\sqrt{s_{\rm NN}}$ =2.76 TeV}",
    eprint = "1401.1250",
    archivePrefix = "arXiv",
    primaryClass = "nucl-ex",
    reportNumber = "CERN-PH-EP-2013-230",
    doi = "10.1016/j.physletb.2014.07.011",
    journal = "Phys. Lett. B",
    volume = "736",
    pages = "196--207",
    year = "2014"
}

@article{ALICE:2016dei,
    author = "Adam, Jaroslav and others",
    collaboration = "ALICE",
    title = "{Multiplicity dependence of charged pion, kaon, and (anti)proton production at large transverse momentum in p-Pb collisions at $\mathbf{\sqrt{{\textit s}_{\rm NN}}}$ = 5.02 TeV}",
    eprint = "1601.03658",
    archivePrefix = "arXiv",
    primaryClass = "nucl-ex",
    reportNumber = "CERN-EP-2016-003",
    doi = "10.1016/j.physletb.2016.07.050",
    journal = "Phys. Lett. B",
    volume = "760",
    pages = "720--735",
    year = "2016"
}

@article{ALICE:2013cdo,
    author = "Abelev, Betty Bezverkhny and others",
    collaboration = "ALICE",
    title = "{$K^0_S$ and $\Lambda$ production in Pb-Pb collisions at $\sqrt{s_{NN}}$ = 2.76 TeV}",
    eprint = "1307.5530",
    archivePrefix = "arXiv",
    primaryClass = "nucl-ex",
    reportNumber = "CERN-PH-EP-2013-132",
    doi = "10.1103/PhysRevLett.111.222301",
    journal = "Phys. Rev. Lett.",
    volume = "111",
    pages = "222301",
    year = "2013"
}

@article{Bierlich:2015rha,
    author = "Bierlich, Christian and Christiansen, Jesper Roy",
    title = "{Effects of color reconnection on hadron flavor observables}",
    eprint = "1507.02091",
    archivePrefix = "arXiv",
    primaryClass = "hep-ph",
    reportNumber = "LU-TP-15-26, MCNET-15-16",
    doi = "10.1103/PhysRevD.92.094010",
    journal = "Phys. Rev. D",
    volume = "92",
    number = "9",
    pages = "094010",
    year = "2015"
}

@article{Werner:2012sv,
    author = "Werner, K.",
    title = "{Lambda-to-Kaon Ratio Enhancement in Heavy Ion Collisions at Several TeV}",
    eprint = "1204.1394",
    archivePrefix = "arXiv",
    primaryClass = "nucl-th",
    doi = "10.1103/PhysRevLett.109.102301",
    journal = "Phys. Rev. Lett.",
    volume = "109",
    pages = "102301",
    year = "2012"
}

@article{Minissale:2015zwa,
    author = "Minissale, Vincenzo and Scardina, Francesco and Greco, Vincenzo",
    title = "{Hadrons from coalescence plus fragmentation in AA collisions at energies available at the BNL Relativistic Heavy Ion Collider to the CERN Large Hadron Collider}",
    eprint = "1502.06213",
    archivePrefix = "arXiv",
    primaryClass = "nucl-th",
    doi = "10.1103/PhysRevC.92.054904",
    journal = "Phys. Rev. C",
    volume = "92",
    number = "5",
    pages = "054904",
    year = "2015"
}

@article{Cho:2019lxb,
    author = "Cho, Sungtae and Sun, Kai-Jia and Ko, Che Ming and Lee, Su Houng and Oh, Yongseok",
    title = "{Charmed hadron production in an improved quark coalescence model}",
    eprint = "1905.09774",
    archivePrefix = "arXiv",
    primaryClass = "nucl-th",
    doi = "10.1103/PhysRevC.101.024909",
    journal = "Phys. Rev. C",
    volume = "101",
    number = "2",
    pages = "024909",
    year = "2020"
}

@article{Begun:2014rsa,
    author = "Begun, Viktor and Florkowski, Wojciech and Rybczynski, Maciej",
    title = "{Transverse-momentum spectra of strange particles produced in Pb+Pb collisions at $\sqrt{s_{\rm NN}}=2.76$ TeV in the chemical non-equilibrium model}",
    eprint = "1405.7252",
    archivePrefix = "arXiv",
    primaryClass = "hep-ph",
    doi = "10.1103/PhysRevC.90.054912",
    journal = "Phys. Rev. C",
    volume = "90",
    number = "5",
    pages = "054912",
    year = "2014"
}

@article{Pierog:2013ria,
    author = "Pierog, T. and Karpenko, Iu. and Katzy, J. M. and Yatsenko, E. and Werner, K.",
    title = "{EPOS LHC: Test of collective hadronization with data measured at the CERN Large Hadron Collider}",
    eprint = "1306.0121",
    archivePrefix = "arXiv",
    primaryClass = "hep-ph",
    reportNumber = "DESY-13-125",
    doi = "10.1103/PhysRevC.92.034906",
    journal = "Phys. Rev. C",
    volume = "92",
    number = "3",
    pages = "034906",
    year = "2015"
}

@article{JETSCAPE:2025wjn,
    author = "Sengupta, A. and others",
    collaboration = "JETSCAPE",
    title = "{Hybrid Hadronization -- A Study of In-Medium Hadronization of Jets}",
    eprint = "2501.16482",
    archivePrefix = "arXiv",
    primaryClass = "hep-ph",
    month = "1",
    year = "2025"
}

\end{document}